\documentclass[aps,floatfix,showpacs]{revtex4}
\usepackage{amsmath}
\usepackage{amssymb}
\usepackage{graphics}
\usepackage{hyperref}

\newcommand{\figwidth}{12.00cm}

\begin{document}

\title{Dark channels in resonant tunneling transport through
  artificial atoms}

\author{Eduardo Vaz}
\author{Jordan Kyriakidis}
\homepage{http://soliton.phys.dal.ca}
\affiliation{Department of Physics and Atmospheric Science, Dalhousie
  University, Halifax, Nova Scotia, Canada, B3H 3J5}

\begin{abstract}
  We investigate sequential tunneling through a multilevel quantum dot
  confining multiple electrons, in the regime where several channels
  are available for transport within the bias window.  By analyzing
  solutions to the master equations of the reduced density matrix, we
  give general conditions on when the presence of a second transport
  channel in the bias window quenches transport through the quantum dot.  
  These conditions are in terms of distinct tunneling anisotropies
  which may aid in explaining the occurrence of negative differential
  conductance in quantum dots in the nonlinear regime.
\end{abstract}

\date{\today}


\maketitle

\section{Introduction}

The properties of electron transport in nanoscopic semiconductor
devices are an important diagnostic tool in the basic study of charge
and spin degrees of freedom of functional and reliable nanoelectronic
devices~\cite{Kouwen_Austing_Tarucha}.  In quantum dots defined
electrostatically in a two dimensional electron
gas~\cite{ashoori96:_elect_in_artif_atoms}, both the electron number
and the tunnel coupling to the leads are tunable.  When high potential
barriers separate the quantum dot from the source and drain
reservoirs, tunneling to and from the dot is weak and the number of
electrons $N$ on the dot is a well-defined integer.  In the Coulomb
blockade (CB) regime~\cite{beenak91:coulom.block}, transport proceeds
primarily by sequential resonant tunneling events and $N$ fluctuates
by one.  In the non-equilibrium regime, where more than one transport
channel lies within the bias window, a classical transport
treatment~\cite{Jacobini_MC} is no longer appropriate, and a
nonequilibrium quantum approach is imperative~\cite{Shah, Scholl}.
In this paper, we microscopically derive analytic expressions for the 
current in the sequential regime through the formalism of the reduced density matrix
and generalized master equations for the evolution of the occupation
probabilities~\cite{Blum_DMT}. One of the main motivations is the recent
observation of negative differential conductance in theory and experiment
~\cite{rogge-2005-72,rogge-2005-NDC,cavaliere04:_negat_differ_conduc,
hettler02:_non_linear_trans, thielmann-2005-71} in several specific systems. 
Our goal in the present work is to derive rather general but well-defined
conditions for the suppression or even the quenching of current whenever a
second channel enters the bias window.

\section{model}

We define our model as a two-dimensional lateral quantum dot~\cite{Qdots,
  weis93:compet.chann.singl} weakly coupled to source and drain
reservoirs, where the total Hamiltonian is given by
\begin{subequations}
\begin{equation}
  \label{eq:mainHamil}
  H = H_S + H_{QD} + H_D + H_T.
\end{equation}
The source and drain Hamiltonians, $H_S$ and $H_D$ respectively, are
taken to be noninteracting Fermion systems shifted by the bias:
\begin{equation}
  \label{eq:HReservoir}
  H_{S(D)} = \sum_{s(d)} \left ( \epsilon_{s(d)} \pm \frac{1}{2} eV_B \right ) 
  d^\dag_{s(d)} d_{s(d)},
\end{equation}
with $d^\dag_{s(d)}$ and $d_{s(d)}$ a creation and an anihilation operator,
respectively, for particles in the source (drain) reservoir.

The quantum dot Hamiltonian in Eq.\eqref{eq:mainHamil} is given by,
\begin{equation}
  \label{eq:QDotHamil}
  H_{QD} = \sum_i (\hbar\omega_i + eV_g) c^\dag_i c_i + V_{\mathrm{int}},
\end{equation}
where the single-particle energies $\hbar
\omega_i$ are shifted by the applied gate voltage $V_g$, and
$V_{\mathrm{int}}$ is the interaction among the confined particles, which we
leave general in the present work.

Finally, the coupling between the quantum dot and reservoirs is described
by the tunneling Hamiltonian,
\begin{equation}
  \label{eq:TunnelHamil}
  H_T = \sum_k \left[ \left( \sum_s T^s_k d_s + \sum_d T^d_k d_d\right)
    c^\dag_k + h.c.\right],
\end{equation}
\end{subequations}
where $h.c.$ denotes the hermitian conjugate, and where $T^{s(d)}_k$ is
the tunneling coefficient for a particle tunneling through the barrier
between the single-particle states $|s\rangle$ ($|d\rangle$) in the
source (drain) reservoir and $|k\rangle$ in the dot.  We assume operators
in each subsystem are independent of each other, and that the eigenstates
of the system are known.  For $V_{\mathrm{int}} \neq 0$ in (\ref{eq:QDotHamil}),
these states are, in general, correlated states---coherent superpositions of Slater
determinants.

\section{quantum master equation}

We derive master equations for a non-equilibrium system ($V_{SD}>0$) by means
of the reduced density matrix, $\rho_{mn}$ of the system~\cite{Blum_DMT}. 
In the Born-Redfield theory~\cite{redfield-1957,toutounji-2005}, the diagonal elements are given by
\begin{equation}
  \label{gme}
  \dot{\rho}_{mm} = \sum_n W_{mn} \rho_{nn} - \sum_n W_{nm} \rho_{mm}.
\end{equation}
The general form of the transition rate $W_{mn}$ from state
$|n\rangle$ to $|m\rangle$, where both states are in general many-body
correlated states, is given by~\cite{vaz06:_quant_statis},
\begin{equation}
  \label{eq:Wmn}
  W_{mn} = \frac{2 \pi}{\hbar} \sum_{\alpha \beta} 
  \int_{-\infty}^{\infty} \! d\omega \,
  \left[
    \tilde{A}_{nm}^{\alpha\beta}(\omega, \epsilon_F - v^-)
    B^S_{\alpha\beta}(\omega + v^-)
    + \tilde{A}_{nm}^{\alpha\beta}(\omega, \epsilon_F - v^+)
    B^D_{\alpha\beta}(\omega + v^+)
  \right],
\end{equation}
where $v^{\pm} = e(V_g \pm V_B/2)$, and where
\begin{subequations}
\begin{gather}
  \label{eq:spectralAll}
  \tilde{A}_{nm}^{\alpha\beta}(\omega, \Omega) = 
  A_{nm}^{\alpha\beta}(\omega) \Theta(\omega - \Omega) +
  A_{mn}^{\alpha\beta}(\omega) \Theta(\Omega - \omega), \\
  \label{eq:spectralReduced}
  A_{nm}^{\alpha\beta}(\omega) = 
  \langle n | c_\alpha^{\dag} | m \rangle 
  \langle m | c_\beta | n \rangle \delta(\omega - \omega_{nm})
\end{gather}
\end{subequations}
is a generalized quantum dot spectral function with $\omega_{nm} =
\omega_n - \omega_m$ and $\Theta(\omega)$ a step function.  The
analogous spectral function for the source reservoir at zero
temperature is
\begin{equation}
  \label{eq:spectralReservoir}
  B^S_{\alpha\beta}(\omega) = \sum_s T^s_{\alpha} T^{s*}_{\beta}
  \delta(\omega - \epsilon_s),
\end{equation}
and $B^D_{\alpha\beta}(\omega) = B^S_{\alpha\beta}(\omega)|_{s
  \rightarrow d}$.

The first term in the integrand of~(\ref{eq:Wmn}) describes the dot
interaction with the source reservoir.  This portion of $W_{mn}$ may
be written as $W_{mn}^{S+} + W^{S-}_{mn}$ respectively denoting the
addition and removal of an electron to or from the dot, corresponding
to the two terms of~(\ref{eq:spectralAll}).  The second term
in~(\ref{eq:Wmn}), describes analogous interactions with the drain
reservoir, allowing us to symbolically decompose~(\ref{eq:Wmn}) into
four terms,
\begin{equation}
  \label{eq:WPieces}
  W_{mn} = W^{S+}_{mn} + W^{S-}_{mn} + W^{D+}_{mn} + W^{D-}_{mn},
\end{equation}
denoting the addition and removal of particles from the source and
drain reservoirs.

It is useful to define a generalized chemical potential $\mu^n_N$, as
the energy required to add a particle to the $N$-particle ground
state, yielding an $(N + 1)$-particle system in the $n^{\mathrm{th}}$
excited state ($n = 0$ is therefore the usual chemical potential).
That is, $\mu^n_N = E^n_{N+1} - E^0_N$, where $E^n_{N+1}$ is the
$n^{\mathrm{th}}$ excited state of the $(N + 1)$-particle system, and
$E^0_N$ is the ground state of the $N$-particle system.  At zero
temperature and zero bias, the system is in the $N$-particle ground
state when the chemical potential $\mu^0_{N-1}$ lies below the Fermi
energy of the reservoirs, and $\mu^0_N$ lies above the Fermi energy.

\section{transition conditions}

If we denote by $N$ the number of confined particles at equilibrium,
and by $N_k$ the number of particles in the state $|k\rangle$, then
by an analysis of Eqns.~(\ref{eq:spectralAll}), (\ref{eq:spectralReduced})
and~(\ref{eq:spectralReservoir}), one can deduce that $W_{mn}$ is non-zero
only when one or more of the following four sets of conditions are satisfied:
\begin{subequations}
\label{eq:Wcond}
\begin{gather}
  \label{eq:wcond1}
  N < N_m = N_n + 1, \ %
  (\nu_N^{N_n} + g^+) < (\Delta_{nm} \pm \frac{1}{2} eV_B),
  \\ \label{eq:wcond2}
  N \geqslant N_m = N_n + 1, \ %
  (\nu_{N_n}^{N-1} + g^-) > (\Delta_{mn} \mp \frac{1}{2} eV_B),
  \\ \label{eq:wcond3}
  N \leqslant N_m = N_n - 1, \ %
  (\nu_{N_m}^N + g^+) > (\Delta_{mn} \pm \frac{1}{2} eV_B),
  \\ \label{eq:wcond4}
  N < N_m = N_n - 1,  \ %
  (\nu_{N_m}^{N-1} + g^-) < (\Delta_{nm} \mp \frac{1}{2} eV_B).
\end{gather}
\end{subequations}
Here, $\nu_N^{M} \equiv \mu^0_{M} - \mu^0_N$ denotes the difference in chemical
potentials between $M$ particles and $N$ particles in the dot.  In addition, $g^+ \equiv
\mu^0_{N} - (\epsilon_F - eV_g)$ denotes the energy difference between the
chemical potential of the $N$-particle dot and the chemical potential of the
source reservoir, and similarly $g^- \equiv (\epsilon_F - eV_g) - \mu^0_{N - 1}$
denotes the energy difference between the chemical potential of the drain
reservoir and that of the $(N-1)$-particle dot.  The energy
difference $\Delta_{mn} \equiv \delta E^m - \delta E^n$, where $\delta E^n$
is the excitation energy of the state $|n\rangle$.  That is, if $E_0^{N_k}$
is the ground-state energy of the $N_k$-particle system, then $E^k = E_0^{N_k}
+ \delta E^k$ with $\delta E^k \geqslant 0$.  The relations $N \lessgtr N_m =
N_n \pm 1$ in Eq.~(\ref{eq:Wcond}) are a consequence of sequential tunneling
by (only) a single electron.  On the other hand, the relations $(\nu^{N_2}_{N_1}
+ g^{\pm}) \lessgtr (\Delta_{pq} \pm eV_B/2)$ are a consequence of resonant
tunneling.  Moving out of the Coulomb blockade regime relaxes the first set
of inequalities while non-Markovian effects relax the second set.

\section{currents: 1 and 2 channel systems}

We consider the current through a quantum dot such that electrons tunnel
into the dot from the source reservoir, and out of the dot to the drain. 
 The evolution of the diagonal elements of the density
matrix~\eqref{gme} for the system depends on both the source and drain reservoirs
and can be separated into a contribution due to coupling with the source, and a
contribution due to coupling with the drain,
\begin{equation}
  \label{gme_SD}
  \dot{\rho}_{mm} = \ \dot{\rho}^S_{mm} + \dot{\rho}^D_{mm}
  = \sum_n\left[ \left( W^{S+}_{mn}\rho_{nn} - W^{S+}_{nm}\rho_{mm}\right) 
    + \left( W^{D-}_{mn}\rho_{nn} - W^{D-}_{nm}\rho_{mm} \right) \right].
\end{equation}

We denote current into the dot through the source
barrier by $I^{S}$, and current out of the dot through the drain barrier by $I^{D}$. 
At long times, the current is proportional to the difference between
these two quantities: $I = I^S - I^D =  e\langle \dot{N} \rangle_{S} - e\langle \dot{N}
\rangle_{D}$, where, for example, $\langle \dot{N} \rangle_S = \sum_m N_m \dot{\rho}^S_m$.
  
Using~(\ref{gme_SD}) we obtain,
\begin{subequations}
\label{eq:current}
\begin{gather}
  \label{eq:sourceCurrent}
  I^S = e \sum_{mn} N_m 
  \left(W^{S+}_{mn} \rho_n - W^{S+}_{nm} \rho_m\right)
  \\ \label{eq:drainCurrent}
  I^D = e \sum_{mn} N_m
  \left(W^{D-}_{nm} \rho_m - W^{D-}_{mn} \rho_n \right),
\end{gather}
\end{subequations}
and $I^S = -I^D$ in the steady-state regime.

\begin{figure}
  \begin{center}
    \resizebox{\figwidth}{!}{\includegraphics{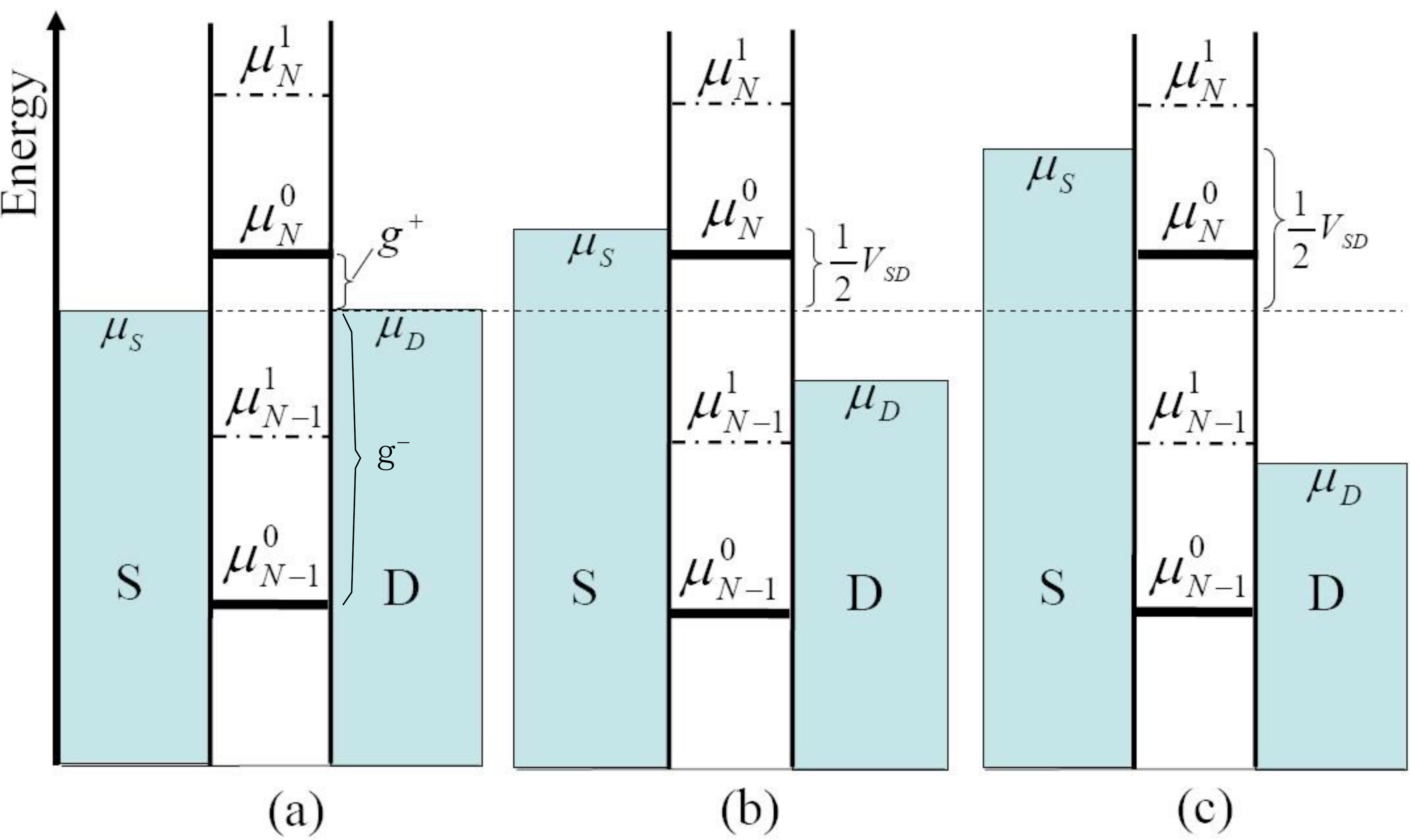}}
    \caption{\label{energ1}Diagram of dot energy regimes.  (a) No
      transport channels in the bias window, (b) single transport
      channel, $\mu^0_N$, in the bias window, (c) two transport
      channels, $\mu^0_N$ and $\mu^1_{N-1}$, in the bias window.}
  \end{center}
\end{figure}

We consider the configuration shown in Fig.~\ref{energ1}, where the
initial equilibrium state of the system is the $N$-particle ground
state, and we calculate the steady state currents through the dot
for the respective cases of one (Fig.~\ref{energ1}b) and two (Fig.~\ref{energ1}c) 
channels in the transport window.  We define three many-body states: 
$|1\rangle \equiv |N\rangle_0$ is the $N$-particle ground state, $|2\rangle \equiv
|N+1\rangle_0$ is the ($N+1$)-particle ground state, and $|3\rangle
\equiv |N\rangle_1$ is the $N$-particle first excited state.  The
transitions between these states are governed in part by
Eq.~(\ref{eq:Wcond}).

There will be precisely one channel in the transport window,
Fig.~\ref{energ1}b, whenever all of the following three inequalities
are satisfied:
\begin{subequations}
  \label{eq:oneChannel}
  \begin{gather}
    \label{eq:oneChannel1}
    g^+ < \frac{1}{2} eV_B < g^-
    \\ \label{eq:oneChannel2}
    \frac{1}{2} eV_B < \Delta^N - g^+
    \\ \label{eq:oneChannel3}
    \frac{1}{2} eV_B < \Delta^{N + 1} + g^+,
  \end{gather}
\end{subequations}
where $\Delta^N \equiv E^N_{\mathrm{1^{st} exc}}-E^N_{\mathrm{ground}}$ 
is the excitation energy of the first excited state of the $N$-particle 
system (this excitation can be due to an electron-electron interaction, or 
due to Zemann spliting in the presence of a magnetic field~\cite{hanson-2004-70}).  
Assuming initially unit occupation of only the $N$-particle ground 
state, the infinite coupled set~(\ref{gme}) reduces, on application 
of~(\ref{eq:Wcond}) to two equations:
\begin{subequations}
\begin{gather}
  \label{eq:MastEqnOneChannel1}
  \dot{\rho}_{11} = W_{12} \rho_{22} - W_{21} \rho_{11}, \\
  \label{eq:MastEqnOneChannel2}
  \dot{\rho}_{22} = W_{21} \rho_{11} - W_{12} \rho_{22}.
\end{gather}
\end{subequations}
That is, transport involves the ground states of the $N$ and
($N+1$)-particle systems (only), as expected.  Solving this coupled
set and inserting into~(\ref{eq:current}), the steady state currents
are found to be
\begin{equation}
  \label{currents_b}
  I_{b} = I_{b}^S = I_{b}^D = e \frac{W^{S+}_{21} W^{D-}_{12}}{W^{S+}_{21} + W^{D-}_{12}}
\end{equation}
which is the standard expression~\cite{meir92:landauer.formul.curren,
  thielmann-2003-68} for a two-level steady state dc current.

The two-channel regime, Fig.~\ref{energ1}c, is entered upon
relaxing condition~(\ref{eq:oneChannel2}).  If we again assume unit
occupation of the $N$-particle ground state as our initial condition,
the set~(\ref{gme}) now reduces to three equations:
\begin{subequations}
  \begin{gather}
    \dot\rho_{11} = W_{12} \rho_{22} - W_{21} \rho_{11}
    \\
    \dot\rho_{22} = W_{21} \rho_{11} + W_{23} \rho_{33} - 
    \rho_{22} (W_{12} + W_{32})
    \\
    \dot\rho_{33} = W_{32} \rho_{22} - W_{23} \rho_{33}
  \end{gather}
\end{subequations}
Transport now involves the ground states of the $N$ and
$(N+1)$-particle ground states---states $|1\rangle$ and $|2\rangle$
respectively---as well as the first excited state with $N$
particles---state $|3\rangle$.  (Note that if
condition~(\ref{eq:oneChannel3}) rather that~(\ref{eq:oneChannel2})
were relaxed, the expressions, and the states involved, would differ.)
Solving this coupled set and inserting into~(\ref{eq:current}), the
steady state currents are found to be
\begin{equation}
  \label{currents_c}
  I_{c} = I_{c}^S = I_{c}^D = \frac{e W^{S+}_{21} W^{S+}_{23} 
    (W^{D-}_{12} + W^{D-}_{32})}{W^{S+}_{21} W^{D-}_{32} + 
    W^{S+}_{23} (W^{D-}_{12} + W^{S+}_{21})}.
\end{equation}

Upon increasing the bias from zero, the steady-state transport current should
change from zero to $I_b$~\eqref{currents_b}, and finally to $I_c$~\eqref{currents_c}. 
Negative differential conductance will be observed whenever $I_{c}<I_{b}$. 
That is, an increase in the bias reduces the current.  General conditions 
under which such phenomena occur are explored in the subsequent sections.

\section{asymmetries}
To explore the relative magnitudes of $I_b$ and $I_c$, we define two
anisotropies.  The first is an extrinsic anisotropy characterizing the
barrier widths.  If we assume tunneling to and from the source is
proportional to tunneling to and from the drain, we can write
\begin{equation}
  \label{eq:extrinsicAnis}
  W^{S+}_{mn} / W^{D-}_{nm} \equiv \alpha_{mn}
\end{equation}
for all many-body states $|n\rangle$ and $|m\rangle$.  Through
Eq.~(\ref{eq:Wmn}), these can be related directly to the tunneling
coefficients $T^d_k$, $T^s_l$ appearing in
Eq.~(\ref{eq:TunnelHamil}).  In particular, the noninteracting limit yields
\begin{equation}
  \label{eq:extrinsicAnisNoninteracting}
  \alpha_{mn} \xrightarrow[\text{limit}]{\text{non-interacting}}
  |T^{s_0}_{k_0}|^2 / |T^{d_0}_{k_0}|^2,
\end{equation}
where $T^{s_0}_{k_0}$ and $T^{d_0}_{k_0}$ are the tunneling
amplitudes of the resonant states.  In general, $0 < \alpha < \infty$.

The second anisotropy is intrinsic to the dot and denotes the relative
ease with which particles can tunnel to different orbitals.  Core
orbitals tightly bound to the center of the dot, for example, have
weaker coupling to the leads than do orbitals localized around the
edge of the dot.  for the particular case of the many-body states
$|1\rangle$, $|2\rangle$, and $|3\rangle$ defined above, we write
\begin{equation}
  \label{eq:intrinsicAnis}
  W^{D-}_{12} / W^{D-}_{32} \equiv \epsilon,
\end{equation}
where $|1\rangle$ and $|3\rangle$ are $N$-body states, and $|2\rangle$
is an $(N+1)$-body state.  The magnitude of $\epsilon$ will be
determined by the spatial extent of the single-particle orbitals
involved in the respective transitions.  Starting from the $(N+1)$-particle
state $|2\rangle$, $W^{D-}_{12}$ describes an electron tunneling out of the dot,
leaving the system in the $N$-particle state $|1\rangle$, whereas $W^{D-}_{32}$
describes a similar process, with the system ending up in state $|3\rangle$.
The ration $\epsilon$ depends on the particular orbitals involved.

In the example we give below, $1 < \epsilon < \infty$.

\section{quenching of current}

\begin{figure}[b]
  \resizebox{\figwidth}{!}{\includegraphics{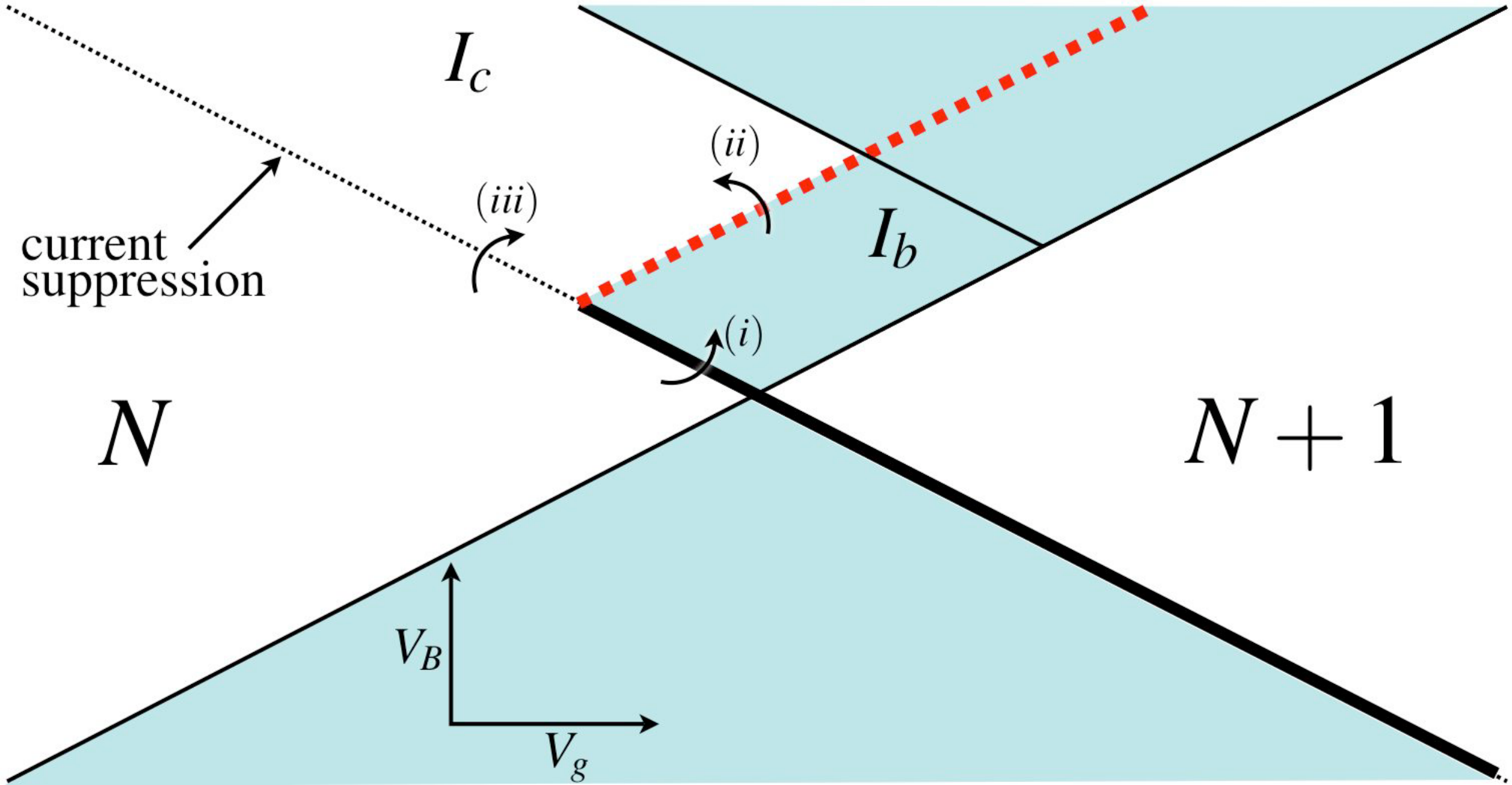}}
  \caption{\label{fig:coulomdDiamond}Coulomb diamond detail showing
    suppression of ground-state current.  the current increases along
    $(i)$, but may decrease along $(ii)$.  In extreme cases, the
    decrease along $(ii)$ matches the increase along $(i)$, yielding
    no current along $(iii)$.  $I_b$ and $I_c$ correspond respectively
    to (b) and (c) of Fig.~\ref{energ1}, and their analytic expressions are given in 
Eqns.~\eqref{currents_b} and~\eqref{currents_c} respectively.}
\end{figure}

Examining the currents $I_c$, Eq.~(\ref{currents_c}), and $I_b$,
Eq.~(\ref{currents_b}), in terms of the anisotropy parameters
$\alpha_{mn}$ and $\epsilon$, we find
\begin{equation}
  \label{eq:currentRatio}
  \frac{I_c}{I_b} = \frac{(1 + \epsilon) (1 + \alpha_{21})}
  {\epsilon \left(1 + \alpha_{21} +
      \frac{\alpha_{21}}{\alpha_{23}}\right)} \equiv R.  
\end{equation}
For $R < 1$, current is suppressed and, in the extreme case of
$\alpha_{23} \rightarrow 0$ (i.e., $W^{S+}_{23} \rightarrow 0$), the
presence of the second channel in the transport window completely
quenches the current, $I_c \rightarrow 0$.  A detail of the Coulomb
diamond for which this analysis applies is given in
Fig.~\ref{fig:coulomdDiamond}.

\begin{figure}[t]
  \resizebox{\figwidth}{!}{\includegraphics{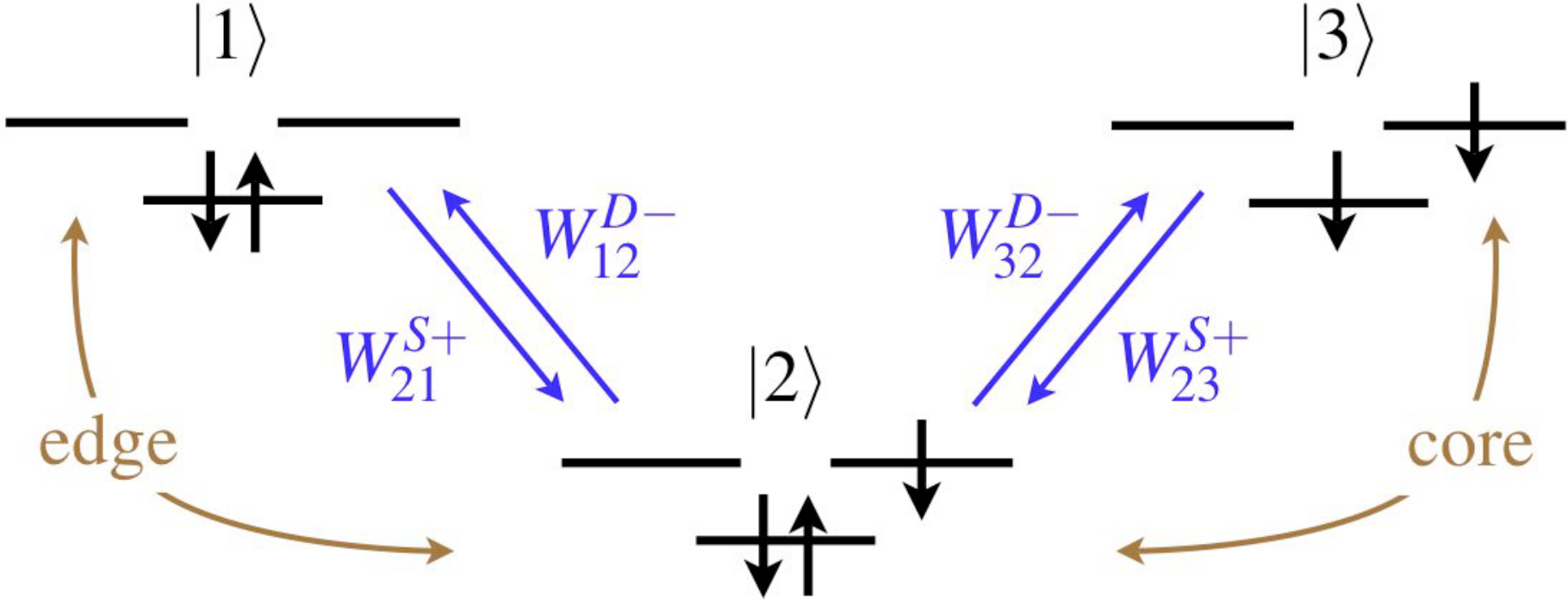}}
  \caption{\label{fig:2eConfig}Configurations capable of quenching
    ground-state transport.  Rates involving states $|1\rangle$ and
    $|2\rangle$ involve tunneling of an electron into a dot
    edge-state (more probable) while those involving states
    $|3\rangle$ and $|2\rangle$ involve tunneling into a dot
    core-state (less probable).}
\end{figure}

As an illustrative example, we consider a two-dimensional parabolic
quantum dot, confining two electrons, with sequential resonant
tunneling proceeding by the addition and removal of a third electron.
The three states corresponding to Fig.~\ref{energ1}c are the
ground-state singlet, $|1\rangle$, the lowest-energy triplet,
$|3\rangle$, and the three-particle ground state, $|2\rangle$.  These
respective configurations are shown in Fig.~\ref{fig:2eConfig}.  The
orbitals can be labeled by two harmonic oscillator indices plus spin,
$|mns\rangle$, where the shell number is given by $m + n = 0, 1, 2,
\ldots$, and the angular momentum $L_z = n - m = 0, \pm 1, \pm 2,
\ldots$.  With regard to Fig.~\ref{fig:2eConfig}, we have $|1\rangle =
|00\uparrow, 00\downarrow\rangle$, $|2\rangle = |00\uparrow,
00\downarrow, 01\downarrow\rangle$, and $|3\rangle = |00\downarrow,
01\downarrow\rangle$.  The orbital $|01\rangle$ penetrates the
barrier~\cite{Kouwen_Austing_Tarucha} to a greater degree than the
$|00\rangle$ orbital, and so the tunneling amplitude $T^{s(d)}_{00}$
has smaller magnitude than $T^{s(d)}_{01}$; depending on the geometry
and the magnetic field, this difference can be severe.  In general, we
expect $\epsilon > 1$ in Eq.~(\ref{eq:intrinsicAnis}).

In the two-channel regime, we look first at the symmetric barrier
case, $\alpha_{mn} = 1$.  Here, suppression of current, $R < 1$,
Eq.~(\ref{eq:currentRatio}), occurs whenever $\epsilon > 2$, that is,
whenever tunneling into the edge is more than two times as likely as
tunneling into the core.  In this case, current is never completely
quenched; the limiting case ($\epsilon \rightarrow \infty$) has $R
\rightarrow 2/3$.

Complete quenching does occur, however, if we additionally have a
barrier asymmetry.  Specifically, if the source barrier is
sufficiently opaque such that tunneling into the core from the source
reservoir is suppressed ($W^{S+}_{23} \rightarrow 0$, see
Fig.~\ref{fig:2eConfig}), but tunneling from the core to the drain is
not ($W^{D-}_{32} \neq 0$), then $\alpha_{23} \rightarrow 0$ and
$I_c \rightarrow 0$ even though $I_b \neq 0$; the addition of the
second transport channel in the bias window suppresses the current
completely.  Once state $|3\rangle$ becomes occupied, the system is
blocked and no other transitions can occur.

This quenching of current is a \emph{dynamical blockade}: it requires an 
interplay between both ground and excited states.  This is in contrast to 
effects such as a) spin blockade, where a transition changing the number 
of electrons by one is accompanied by a change of spin greater than
1/2~\cite{huttel-epl-2003}, or b) blockade due to spin polarized injection 
and extraction of charge carriers~\cite{kyriakidis-2002-66}, where the current
is quenched due to the Pauli exclusion principle.

The quenching of current will be destroyed whenever the spin-flip plus
orbital decay rate is large compared to the dwell time of the
tunneling electron.  In this case, a direct transition from
$|3\rangle$ to $|1\rangle$ may occur by emission of a phonon and
simultaneous absorption of spin through nuclear baths, spin-orbit
coupling, or other spin-symmetry-breaking process.  
Likewise, strong Coulomb correlations will generally open up
additional tunneling pathways, likely involving edge orbitals, and
this too will destroy the quenching of current.

\section{conclusions}

In conclusion, we have derived a set of general yet well-defined conditions for 
quenching of transport through a multichannel quantum dot.
We find that under certain tunneling anisotropies the presence of a second transport
channel may decrease the current through the dot. Specifically, if the tunneling rates
to or from core states are less than half than the rates to or from edge states, the
current may be suppressed.  Complete quenching of the current may occur if, in addition to the edge/core anisotropy, an anisotropy between source and drain barriers is also
present.  The presence of these dark channels is a measure of the lifetime of excited
states.  

Finally, although the present work focused on the two-electron case,
the general phenomena should appear in numerous locations throughout the
Coulomb diamonds; it can generally appear whenever ground-state tunneling
transitions involve edge orbitals, and excited-states tunneling involves
core orbitals, and asymmetric source and drain barriers are present.

\begin{acknowledgments}
  This work was supported by NSERC of Canada and the Canadian
  Foundation for Innovation.  JK additionally acknowledges
  discussions with the Institute for Microstructural Sciences of NRC
  Canada.
\end{acknowledgments}


\end{document}